\documentclass[preprint,aps,12pt,preprintnumbers,eqsecnum,nofootinbib]{revtex4}
\usepackage{xcolor}
\usepackage{graphicx}
\usepackage{epstopdf} 
\usepackage{braket}
\usepackage{amssymb,amsmath,amsfonts,comment,latexsym,graphicx,epstopdf,float}
\usepackage{color}
\usepackage{fancyvrb}
\usepackage{chngcntr}
\usepackage{etoolbox}
\usepackage{ulem}
\usepackage{array}
\usepackage{slashed}

\usepackage[colorlinks=true,citecolor=blue,linkcolor=red,filecolor=cyan,urlcolor=magenta]{hyperref}

\usepackage{float}

\newcommand{\be}{\begin{equation}}
\newcommand{\ee}{\end{equation}}
\newcommand{\ba}{\begin{eqnarray}}
\newcommand{\ea}{\end{eqnarray}}

\newcommand{\grts}{\raise.3ex\hbox{$>$\kern-.75em\lower1ex\hbox{$\sim$}}}
\newcommand{\lets}{\raise.3ex\hbox{$<$\kern-.75em\lower1ex\hbox{$\sim$}}}

{\catcode`\|=\active\gdef\Braket#1{\left<\mathcode`\|"8000\let|\bravert 
{#1}\right>}}

\def\bravert{\egroup\,\vrule\,\bgroup}

\usepackage{color}
\usepackage{amssymb,amsmath,amsfonts}
\usepackage{epstopdf} 
\usepackage{braket}
\usepackage{bookmark}

\usepackage{autobreak}

\begin{document}
%
%  
% Title of paper
\title{\vspace*{0.5in} 
Monoenergetic Neutrinos from WIMP Annihilation in Jupiter
\vskip 0.1in}
\author{George M. French}\email[]{gmfrench@email.wm.edu}
\author{Marc Sher}\email[]{mtsher@wm.edu}

%`
\affiliation{High Energy Theory Group, Department of Physics,
William \& Mary, Williamsburg, VA 23187-8795, USA} 
%
%  
%%\date{\today}
\date{\today}
\begin{abstract}
Weakly interacting massive particles (WIMPs) can be captured by the Sun and annihilate in the core, which may result in production of kaons that can decay at rest into monoenergetic 236 MeV neutrinos.   Several studies of detection of these neutrinos  at DUNE have been carried out.   It has been shown that if the WIMP mass is below 4 GeV, then they will evaporate prior to annihilation, suppressing the signal.    Since Jupiter has a cooler core, WIMPs with masses in the 1-4 GeV range will not evaporate and can thus annihilate into kaons which decay at rest into monoenergetic neutrinos.    We calculate the flux of these neutrinos near the surface of Jupiter and find that it is comparable to the flux of neutrinos from the Sun at DUNE for masses above 4 GeV and substantially greater in the 1-4 GeV range.    Of course, detecting these neutrinos would require a neutrino detector near Jupiter.   Obviously, it will be many decades before such a detector can be built, but should direct detection experiments find a WIMP with a mass in the 1-4 GeV range, it may be one of the few ways to learn about the annihilation process.   A liquid hydrogen time projection chamber might be able to get precise directional information and energy of these neutrinos (and hydrogen is plentiful in the vicinity of Jupiter).    We speculate that such a detector could be placed on the far side of one of the tidally locked Amalthean moons; the moon itself would provide substantial background shielding and the surface would allow easier deployment of solar panels for power generation.

\end{abstract}

\maketitle

\section{Introduction} \label{sec:intro}

Weakly interacting massive particles (WIMPs) are one of the main candidates for dark matter.    The primary detection strategies for detection of WIMPs are production at colliders, direct detection in underground experiments and indirect detection from WIMP annihilation.     The efficacy of each of these strategies is very dependent on the mass and interactions of the WIMPs, and thus all three must be deployed.    It has been noted \cite{Silk:1985ax,Srednicki:1986vj,Press:1985ug,Krauss:1985ks,Gould:1987ju,Gould:1987ww,Belotsky:2001ka,Belotsky:2002sv} that WIMPs can be gravitationally captured by the Sun, resulting in a much higher WIMP density in the Sun, leading to annihilation into neutrinos (most other annihilation products will not be detectable outside the Sun).     Searches for high energy neutrinos from WIMP annihilation in the Sun have been carried out \cite{IceCube:2021xzo,Super-Kamiokande:2015xms,ANTARES:2016xuh}.     It was later pointed out \cite{Rott:2012qb,Bernal:2012qh} that in models in which the WIMPs annihilate into light quarks (or heavy quarks which then decay into light quarks) there will be a large number of low-energy (sub-GeV) neutrinos produced.    These papers focused on decays of muons and pions.    However, in a series of papers by Rott, In, Kumar and Yaylali (RIKY) \cite{Rott:2015nma,Rott:2016mzs,Rott:2017weo}, it was argued that the pions and kaons would come to rest before decaying and thus would decay into monoenergetic neutrinos.   Pions yield $32$ MeV neutrinos and kaons yield ($64\%$ of the time)  $236$ MeV neutrinos.  RIKY noted that WIMPs with masses below 3-4 GeV would evaporate, but that masses in the $4-10$ GeV range would cover a region of  parameter-space which could be detected at DUNE and would not be excluded by direct detection experiments. A flux of $236$ MeV neutrinos coming from the Sun would be a smoking gun for dark matter annihilation.   Recently, DUNE \cite{DUNE:2021gbm} has analyzed this possibility and shown that spin-dependent cross-sections as low as $10^{-38} {\rm cm}^2$ can be reached.

Detection of a monoenergetic flux of neutrinos from the Sun would certainly tell us a great deal about WIMP dark matter, but unless one also had direct detection or collider evidence, there would remain many unanswered questions.     Are there other celestial bodies that could provide information about dark matter annihilation?  WIMP capture in the Earth would be very rare, since Earth has a much smaller size and a much smaller escape velocity.    In early papers, Kawasaki et al. \cite{Kawasaki:1991eu} and Adler \cite{Adler:2008ky} discussed strongly interacting dark matter as source for heating of gas giant planets,   Leane et al. \cite{Leane:2021ihh} looked at the possibility that dark matter could be focused by celestial bodies, increasing the rate of annihilation and Leane and Linden \cite{Leane:2021tjj} studied gamma ray emission from dark matter annihilation in Jupiter.    Very recently, Li and Fan \cite{Li:2022wix} discussed WIMP capture in Jupiter.    They also pointed out that Jupiter is a particularly promising celestial object because it is the largest gas giant and its core is relatively cool, reducing the evaporation rate.   As a result, WIMPs with masses below the 4 GeV evaporation limit from the Sun could collect in the core.     Li and Fan studied the possibility that the WIMPs could annihilate into long-lived dark mediators which would convert to electrons and positrons after leaving Jupiter.     Current data from the Galileo and Juno orbiters gave interesting constraints on dark matter models.

These works all considered WIMPs that eventually decay into charged particles.     Could one detect a monoenergetic flux of neutrinos from Jupiter?      Obviously, there is no current detector in orbit that could detect neutrinos, nor is there likely to be for many decades.    But such a detector could encounter a huge flux of neutrinos.    The inverse square law alone would give an enhancement of the square of 1 A.U./$R_{\rm Jupiter}$, which is a factor of five million relative to DUNE.    This could far exceed the reduction due to the smaller size (relative to the Sun) of Jupiter and the smaller escape velocity.   Hopefully by the end of this century robust exploration of the Jovian system will be underway and the idea of orbiting a neutrino detector will not be unthinkable.    Obviously if dark matter is detected and annihilation into light quarks is possible, then this type of detector could be helpful.   Even if the annihilation into light quarks is detected at an earlier stage, such a detector could give us direct information about the Jovian interior.    Thus, we feel that it is valuable to study the question of WIMP annihilation into kaons in Jupiter and the detection of the neutrinos, acknowledging that such a detection would be decades away.   

\section{WIMP annihilation in Jupiter} \label{sec:two}
\subsection{WIMP population}
As WIMPs from the DM halo pass through Jupiter, a portion of them scatter off of atomic nuclei and enter into bound orbits. While some scatter back out after additional collisions, the rest remain bound and thermalize in the planet's core where they annihilate into Standard Model particles \cite{Baum:2016oow, Rott:2015nma}. The rate at which the total population of WIMPs changes with time inside of Jupiter is governed by the differential equation
\begin{equation}
\frac{dN_\chi(t)}{dt}= \mathcal{C} - \mathcal{E}N_\chi(t) - \mathcal{A}N^2_\chi(t)
\end{equation}
where the coefficients $\mathcal{C}$, $\mathcal{E}$, $\mathcal{A}$ correspond to capture, evaporation, and annihilation, respectively and $N_\chi$ is the number of WIMPS in Jupiter \cite{Garani:2021feo}. It is assumed throughout this work that the coefficients are time-independent and that WIMPs are their own anti-particle\footnote{It has been pointed out \cite{Rott:2015nma} that if the WIMP is a Majorana fermion and if one assumes minimal flavor violation, then the annihilation rate is suppressed by light quark masses.   Of course, the WIMP could be a scalar or minimal flavor violation might not be realized.  In any event, this assumption will not affect our results substantially.}. The solution to this equation is
\begin{equation}
N_\chi(t) = \frac{\mathcal{C}\tanh(t/\tau)}{\tau^{-1} + (\mathcal{E}/2)\tanh(t/\tau)}
\end{equation}
where $\tau=1/\sqrt{\mathcal{CA} + \mathcal{E}^2/4}$ is the time it takes for the system to reach equilibrium \cite{Liang:2016yjf}. For a 1 GeV WIMP, Li and Fan \cite{Li:2022wix} find that $t_J/\tau \sim 10$ where $t_J \simeq 4 \text{ Gyr}$ is a proxy for the age of the solar system. This is an important result because the annihilation rate, $\Gamma_A = \mathcal{A}N_\chi^2/2$, is maximum when $\tanh(t/\tau) \simeq 1$. Since the outgoing neutrino flux is proportional to the annihilation rate, it is maximized as well.

The full derivations of the capture, annihilation, and evaporation rates are not shown here as they have been done many times. Instead, only brief overviews are given and those interested in a more involved treatment of the derivations are directed to the discussions in previous work \cite{Gould:1987ju,Gould:1987ww,Li:2022wix,Baum:2016oow,Bramante:2017xlb,Garani:2021feo,Kappl:2011kz}.
As a proof of concept, this work only aims for a rough comparison of the monoenergetic neutrino flux near Jupiter to that at a detector on Earth.   For this reason, certain simplifying assumptions will be made consistently throughout this analysis:
\begin{enumerate}
    \item Both Jupiter and the Sun are treated as targets of uniform density.
    \item Jupiter and the Sun are treated as purely hydrogen targets in order to focus on probing the spin-dependent (SD) WIMP-proton cross section $\sigma_p$.   While the results will also apply to the spin-independent  (SI) cross section, direct detection experiments will generally provide much tighter bounds on the SI cross section.
    \item The SD cross section is small enough that Jupiter and the Sun can be treated as optically thin, so only the single-scattering case is considered.
\end{enumerate}

\subsection{Capture}

%%%%%%%%%%%%%%%%%%%%Figure  1  begin %%%%%%%%%%%%%%%%%%%%%%%
\begin{figure}[!htb]
\centering
\includegraphics[width=11cm]{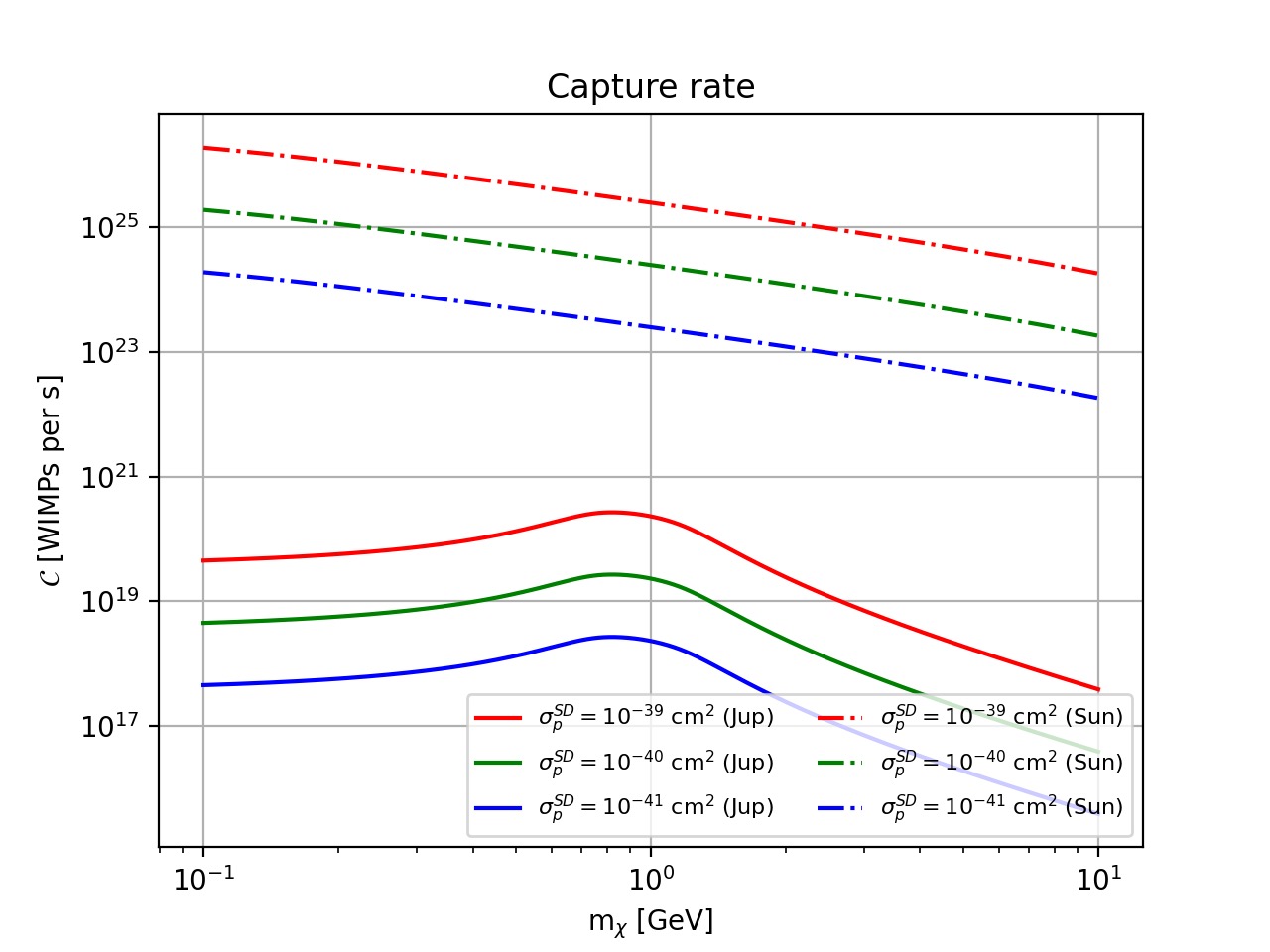}
\caption{\label{cap} 
Capture rate of WIMPs in both Jupiter and the Sun. The resonance peak at roughly $1\text{ GeV}$ corresponds to the WIMP mass being closely matched to the target. This effect is hardly noticeable in the Sun because of its larger escape velocity ($v_{\rm esc,S}\simeq10v_{\rm esc,J}$) \cite{Gould:1987ju,Catena:2016kro}.}
\end{figure}
%%%%%%%%%%%%%%%%%%%%Figure 1 end %%%%%%%%%%%%%%%%%%%%%%%
A flux of WIMPs passing through a thin shell of material will have a certain probability of scattering to a velocity below the escape velocity, $v<v_{\rm esc}$. By integrating over the total volume, one arrives at the capture rate. For an isotropic Boltzmann WIMP distribution with dispersion velocity $v_\chi$, the capture rate is \cite{Gould:1987ju,Baum:2016oow}
\begin{equation}
\mathcal{C} \gtrsim 0.28 \sqrt{\frac{8\pi}{3}} \frac{n_\chi\tau_JR^2_Jv^2_J(R_J)}{\bar{v}_\chi}\bigg(1 - \frac{1 - e^{-A^2}}{A^2}\bigg)
\end{equation}
where $n_\chi$ is the local DM number density, $\tau_J$ is the optical depth given by $\frac{3\sigma_{\chi \rm n}}{2\sigma_{\rm sat}}$ where $\sigma_{\rm sat}$ is the cross section that saturates the geometric limit ($\sigma_{\rm sat,J} = 10^{-34}\ {\rm cm}^2$ and $\sigma_{\rm sat,S} = 10^{-35}\ {\rm cm}^2$) and $\sigma_{\chi \rm n}$ is the DM-nucleon cross section, $v_J(R)$ is the escape velocity a distance $R$ from the center, $\bar{v}_\chi \simeq 270$ km/sec is the DM velocity dispersion and $A(r)^2 \equiv 6 v_J(r)^2m_n m_\chi/[\bar{v}^2_\chi(m_n - m_\chi)^2]$.
We adopt the lower bound for $\mathcal{C}$ for a conservative estimate.  

Our results for the capture rate are given in Figure 1.     Not surprisingly, the capture rate for Jupiter is substantially lower than the Sun due to its smaller size and smaller escape velocity.      The resonance peak at 1 GeV corresponds to the WIMP mass being closely matched to the nucleon mass.   This effect is not noticeable in the Sun because of its larger escape velocity \cite{Gould:1987ju,Catena:2016kro}.

\subsection{Annihilation}
%%%%%%%%%%%%%%%%%%%%Figure  2  begin %%%%%%%%%%%%%%%%%%%%%%%
\begin{figure}[!htb]
\centering
\includegraphics[width=11cm]{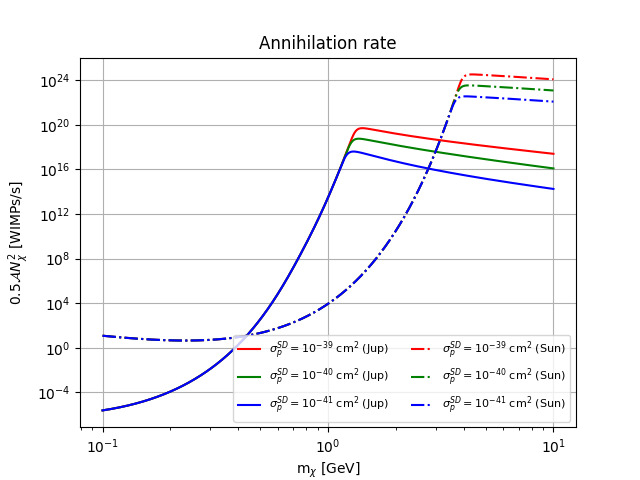}
\caption{\label{ann} 
Annihilation of WIMPs into SM particles in Jupiter and the Sun. It should be noted that this is the total number of WIMP annihilations, but only a fraction of them produce kaons which decay into muon neutrinos. The annihilation rate is comparable to the capture rate above the evaporation mass because both Jupiter and the Sun are in equilibrium ($\tanh(t/\tau)\simeq1$).}
\end{figure}
%%%%%%%%%%%%%%%%%%%%Figure 2 end %%%%%%%%%%%%%%%%%%%%%%%

As the bound WIMPs continue to lose energy in further scatterings and settle in the core, they occupy a region given by a scale radius $r_\chi$. We take this region to be isothermal with temperature $T_c$ and density $\rho_c$. For both the Sun and Jupiter, this can be well approximated by \cite{Garani:2017jcj,Baum:2016oow,Li:2022wix}
\begin{equation}
r_\chi = \sqrt{\frac{3T_c}{2\pi G \rho_c m_\chi}}\simeq0.1R\sqrt{\frac{1\text{ GeV}}{m_\chi}}
\end{equation}
using $T_c = 1.5\times10^4\text{ K}$ ($1.5\times10^7\text{ K}$) and $\rho_c = 2\times10^4\text{ kg m}^{-3}$ ($1.5\times10^5\text{ kg m}^{-3}$) for Jupiter (the Sun). The annihilation coefficient is then \cite{Garani:2021feo,Li:2022wix,Baum:2016oow}
\begin{equation}
\mathcal{A} = \frac{\langle{\sigma v}\rangle_{\rm ann}} {V_{\text{eff}}}
\end{equation}
where $\langle{\sigma v}\rangle_{\rm ann} \sim 10^{-26}\text{ cm}^3\text{/s}$ \cite{Garani:2017jcj,Catena:2016kro} is the thermally averaged annihilation cross section and the effective volume $V_{\rm eff}=4/3\pi r^3_\chi$. In principle, $\langle{\sigma v}\rangle_{\rm ann}$ will vary with $m_\chi$, but we neglect this for the sake of illustration.

The results for the annihilation rate are given in Figure 2.   The rate has been multiplied by the square of the population of WIMPs.    Thus, it includes the effects of evaporation, discussed in the next subsection. This explains the drop one sees in the solar annihilation rate at 4 GeV and in the Jovian annihilation rate at 1 GeV.   Note that only a fraction of annihilations produce kaons.     Above the evaporation mass, the annihilation rate is comparable to the capture rate since Jupiter and the Sun are in equilibrium. 

\subsection{Evaporation}
%%%%%%%%%%%%%%%%%%%%Figure  3  begin %%%%%%%%%%%%%%%%%%%%%%%

\begin{figure}[!htb]
\centering
\includegraphics[width=11cm]{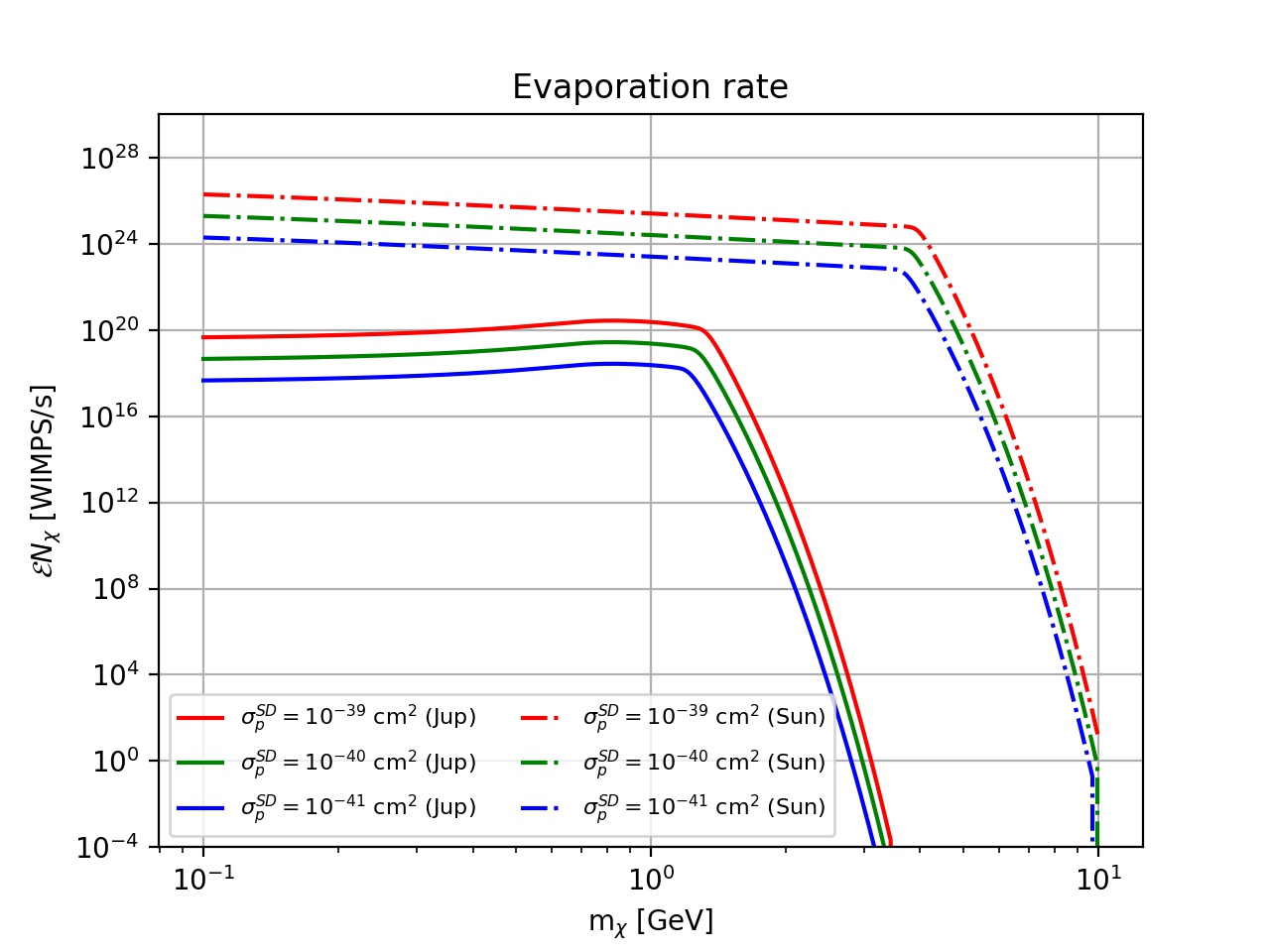}
\caption{\label{evap} 
Evaporation of WIMPs inside Jupiter and the Sun. In both cases, we note that just above a certain mass $m_{\text{evap}}$ evaporation drops sharply to zero, whereas below $m_{\text{evap}}$ it is comparable to the rate of WIMPs being captured. We have estimated this mass to be about 1.3 GeV for Jupiter and 3.3 GeV for the Sun, in agreement with previous estimates \cite{Griest:1986yu,Gould:1987ww,Baum:2016oow,Garani:2021feo,Li:2022wix}. This is not a great restriction on Jupiter as sub-GeV WIMPs are kinematically unable to annihilate into kaons.}
\end{figure}

%%%%%%%%%%%%%%%%%%%%Figure 3 end %%%%%%%%%%%%%%%%%%%%%%%
Similar to the capture rate, calculating the evaporation rate involves taking the probability of a flux of WIMPs scattering to $v>v_{esc}$ for a thin shell of material and integrating over the total volume \cite{Gould:1987ww}. While the full expression is far from transparent, it simplifies a great deal in the limits of $m_\chi \sim m_p$, $E_c = m_\chi v_{\rm esc}^2(0)/2\gg T_c$, and $T_\chi\simeq0.9T_c$ which hold for an order-of-magnitude estimate. Considering the isothermal region $V_{\text{eff}}$ from above, the evaporation coefficient can be approximated by \cite{Gould:1987ww,Garani:2021feo,Catena:2016kro}
\begin{equation}
\mathcal{E} = \sigma_p\frac{N_{0.95}}{V_{\text{eff}}}\bigg(\frac{8T_\chi}{\pi m_\chi}\bigg)^{1/2}\bigg(\frac{E_c}{T_\chi}\bigg)e^{-E_c/T_\chi}
\end{equation}
with $N_{\text{0.95}}$ being the number of protons within the region where $T=0.95T_\chi$. Only a portion of the interior of Jupiter (the Sun) is considered because it corresponds to a region where evaporation is significantly enhanced by the closely matched WIMP and nucleon temperatures. For our purposes, this provides a decent approximation for the overall evaporation rate. We take $N_{0.95} \sim 0.1M/m_p$ which is known to be a reasonable approximation for the Sun \cite{Garani:2021feo,Gould:1987ww,Griest:1986yu}. Here, we also use $v^2_{\text{esc}}(0)\simeq 1.5v^2_{\text{esc}}(R)$ for Jupiter and $v^2_{\text{esc}}(0)\simeq 5v^2_{\text{esc}}(R)$ for the Sun.

The results are plotted in Figure 3.    One sees that the evaporation is negligible for WIMP masses above 3.3 GeV for the Sun and above 1.3 GeV for Jupiter.

\subsection{Neutrino flux}
WIMP annihilations (which occur at a rate $\Gamma_A=\mathcal{A}N^2_\chi/2$) can produce kaons for WIMPs with $m_\chi\gtrsim1\text{ GeV}$. The kaons, upon coming to rest, decay into muon neutrinos via $K^+ \rightarrow \nu_\mu \mu^+$ with a branching ratio of about 64\%. The outgoing flux is given by \cite{Baum:2016oow,Rott:2015nma,Catena:2016kro}
\begin{equation}
\frac{d^2\Phi_{\nu_\mu}}{dEd\Omega} = \frac{\Gamma_A}{4\pi D^2}N_{K}\mathcal{B}_{\nu_\mu}\delta(E-E_0)\delta(\Omega)
\end{equation}
where $D$ is the core-detector distance, $N_K$ is the average number of $K^+$ produced per annihilation, and $\mathcal{B}_{\nu_\mu}$ is the fraction of $K^+$ that decay into $\nu_\mu$. The two dirac delta terms enforce the conditions that the energy signal is monoenergetic ($E_0\approx236\text{ MeV}$) and that all neutrinos emanate from the jovian or solar core, respectively. We can express $N_K$ in terms of the fraction $r_K$ of the c.o.m. energy that is converted into $K^+$ as
\begin{equation}
N_K = \frac{2m_\chi}{m_K}r_K.
\end{equation}
  We take $r_K \sim 1/50 $ for simplicity \cite{Rott:2015nma}. As mentioned above, the condition of equilibrium is important because it means that the flux is maximized for $t\gg\tau$. However, in the region where evaporation dominates, the flux decreases drastically because annihilations occur far more infrequently.   Note that the value of $r_K$ is somewhat uncertain, and should these neutrinos be detected from the Sun, this will pin down the value and be relevant for detection in Jovian orbit.

\section{Detection in Jovian orbit}
%%%%%%%%%%%%%%%%%%%%Figure 4 begin%%%%%%%%%%%%%%%%%%%%%%%
\begin{figure}[!htb]
\centering
\includegraphics[width=11cm]{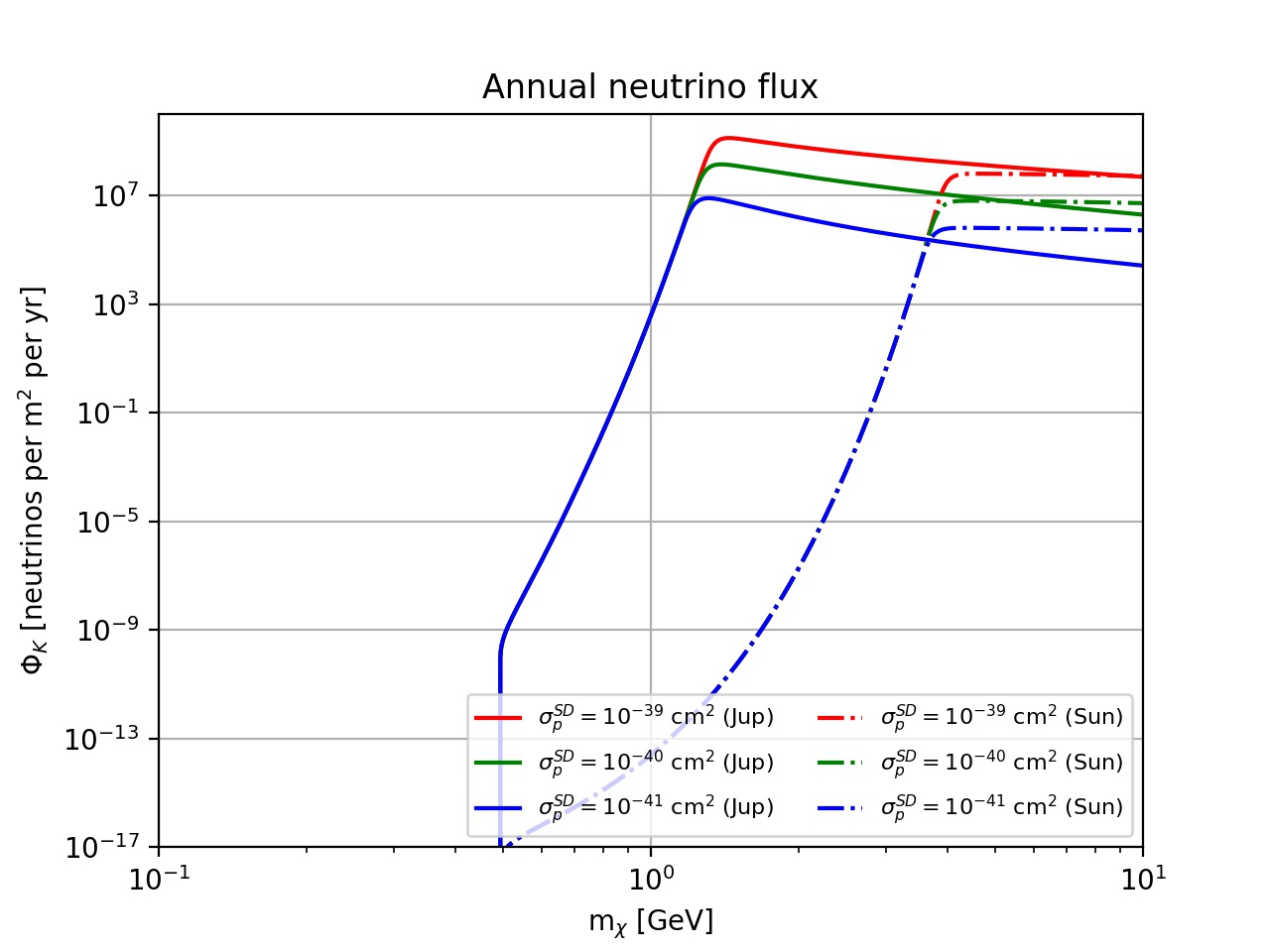}
\caption{\label{flux} 
The dot-dashed lines give the flux of 236 MeV neutrinos at the surface of the Earth from WIMP annihilation in the Sun for three different spin-dependent cross sections.    The solid lines give the flux from WIMP annihilation in Jupiter, near the surface of Jupiter.    Note that the flux near Jupiter is substantially higher in the 1-4 GeV region.   We have  included the phase space factor of $\sqrt{(1-m_K^2/m_{WIMP}^2)}$ in the figure - this is negligible above 1 GeV.}
\end{figure}
%%%%%%%%%%%%%%%%%%%% Figure 4 end %%%%%%%%%%%%%%%%%%%%%%%

From Figure 4, one can see that the flux of neutrinos in low-Jovian orbit is comparable to the flux from the Sun at 1 AU (i.e. at DUNE) in the mass range at or above 4 GeV.    However, in the 1-4 GeV mass range, the flux at Earth orbit is negligible where as the flux in low-Jovian orbit is substantial  (below 1 GeV, WIMP annihilation into kaons becomes negligible due to phase space).    Thus we focus on the $1-4$ GeV mass range.   

If the cross section is spin-dependent, direct detection experiments currently can't detect WIMPs below 2.3 GeV for any cross section.    One reason for this is the energy thresholds of direct-detection experiments. For example, PICASSO reports \cite{PICASSO:2012ngj} a sensitivity to nucleus recoil energies as low as 1.7 keV.       For WIMPs in the halo, this gives a lower bound of 2.3 GeV.    However, a proposal by CYGNUS \cite{Vahsen:2020pzb} could eventually lead to a sensitivity corresponding to WIMP masses as low as 1 GeV.   Thus, in the coming decades, the range of $1-4$ GeV will be explored if the spin-dependent cross section is sufficiently large.     If there is a WIMP in this range, 236 MeV neutrinos from the Sun will not be detectable, and one way to study the annihilation would be to look for neutrinos from Jupiter.

Of course, to detect neutrinos in low-Jovian orbit, one would need to orbit a neutrino detector.   This seems absurd, and for the next few decades is certainly utterly infeasible.   One can imagine that later in this century, there will be human exploration of the Jovian system and a reasonably sized neutrino detector might be thinkable.    In this section the nature of such a detector and its location will be discussed.   We recognize that this is an extremely preliminary discussion, given the unknown nature of technological advances between now and then.    But it is still interesting to speculate.

When a $236$ MeV neutrino interacts with an oxygen or argon nucleus, the charged lepton that emerges is close to isotropic - the Fermi motion of the struck nucleon in the nucleus alone will tend to isotropize the charged lepton.   The proton that emerges, however, will tend to be in the forward direction. Protons with a kinetic energy of a few hundred MeV will not emit Cerenkov radiation and thus water Cerenkov detectors will not be useful (this is unfortunate since the Jovian system has a substantial amount of water/ice).    Liquid argon time projection chamber (TPC) detectors like DUNE would be able to detect these protons and can thus reconstruct the direction and energy of the incident neutrino. Of course, one would need a substantial number of events to determine the average incident neutrino detection - a precise energy and direction determination on an event-by-event basis would not be possible.

A more promising possibility\footnote{We thank Mike Kordosky for this suggestion.} is a liquid hydrogen TPC or bubble chamber detector. There is no Fermi motion and the charged lepton can be easily seen. The neutron emerging from the interaction will travel some distance and interact - that interaction can also be seen. Thus the entire event can be seen, leading to an event-by-event determination of the energy and direction of the initial neutrino. It should be noted that the source isn't at the precise center of Jupiter but typically within $\rm 0.1R_J$, thus the angle of approach will not be completely determined in advance. Neutral current neutrino interactions can also be studied, although the backgrounds might be substantial. One can also note that the lower energy neutrinos from pion decay ($30$ MeV) might be detectable as well, although the length of the nucleon track might be too short. While a large liquid hydrogen detector would be too dangerous to be built on Earth, this would not be a problem in Jovian orbit. And since Jupiter is almost entirely hydrogen, the liquid hydrogen needed for the detector would not have to be transported from Earth (unlike liquid argon or liquid scintillator).

What about the location? Even with good directional information, there will be backgrounds from cosmic ray interactions in the atmosphere. In addition,  power generation would be a more serious problem in low-Jovian orbit - large solar panels attached to the orbiting detector could cause instabilities. One way to avoid these problems would be to place the detector on the back side of one of the tidally-locked Amalthean moons. Solar panels could be spread on the surface fairly easily and the moon itself would provide shielding. Liquid hydrogen would need thermal isolation from sunlight but shielding should not pose great difficulties.

\section{Conclusions}

WIMPs in the galactic halo can interact in the Sun and their velocities can drop below the escape velocity.   These captured WIMPs will gradually fall into the core and annihilate.   While the annihilation products are very model-dependent, there are many models in which they decay into light quarks, leading to production of pions and kaons.  The kaons will quickly slow down in the dense core and decay into (64\% of the time) monoenergetic 236 MeV neutrinos.    Studies have been done calculating the flux of these on Earth and if the WIMP spin-dependent cross section is sufficiently large then the flux will be large enough that these neutrinos can be detected by DUNE.    However, if the WIMP mass is below 4  GeV, then the WIMPs will evaporate before annihilation.   It has been pointed out that Jupiter has a colder core than the Sun and thus WIMPs in the 1-4 GeV range will annihilate.   A study involving annihilation to charged particles via a long-lived mediator was recently carried out.

In this paper, we have calculated the flux of 236 MeV neutrinos from Jupiter near the Jovian ``surface".     Comparing with DUNE, the flux near the surface gets a huge enhancement from the inverse-square law.   For WIMP masses in the 4-10 GeV range, the flux is comparable to DUNE.   However, in the 1-4 GeV range, the flux from the Sun drops off rapidly and the flux near the surface of Jupiter does not.   We studied the possibility of a neutrino detector orbiting Jupiter.    While obviously many decades away, we speculate on the type and location of such a detector.   A liquid hydrogen TPC would possibly allow determination of the energy and direction of 236 MeV neutrinos on an event-by-event basis.    Locating the detector on the far side of a  tidally locked moon orbiting near Jupiter would allow the moon to act as shielding and provide more ease in producing power for the detector.

In the coming decade or two, direct detection searches will cover the entire 1-4 GeV region of WIMP masses down to some spin-dependent cross section level.   Over the decades, this level will drop.   Should a positive signal be found, we will only know the cross-section and mass of the WIMPs.   At that point, one might take the concept of a Jovian neutrino detector seriously as the only way to learn about the annihilation process.

It would be interesting to consider the necessary size of a detector on an Amalthean moon.   As we can see in Figure 4, the flux in the 1-2 GeV mass region is an order of magnitude or two higher than at DUNE.   In addition,  the Amalthean moons orbit at roughly $2 R_{\rm Jupiter}$, leading to a factor of four reduction, so the net flux is roughly an order of magnitude higher.    DUNE is a 34 kiloton LAr TPC, thus a  detector somewhat smaller in mass would give a similar number of events.  However liquid argon has a density roughly 10 times that of liquid hydrogen, thus the physical size of the liquid hydrogen detector would be comparable to DUNE.    The cross section for $236$ MeV neutrinos on a given mass of liquid argon is not dissimilar to that of liquid hydrogen; as pointed out by Rott, et al.\cite{Rott:2017weo}, the energy of the neutrinos is high enough that the impulse approximation is reasonable, so the number of nucleons is relevant.     It would thus seem that a detector size comparable to DUNE would suffice.   As noted above, however, the fact that there is no Fermi motion in hydrogen implies that the entire event can be seen, leading to a substantial reduction in background.    This could significantly reduce the necessary size.   Of course, technological improvements over the next few decades could reduce the size even further. 

Could DUNE itself put bounds on neutrinos from Jupiter?    The flux at DUNE would be smaller by a factor of the square of the ratio of the Earth-Jupiter distance to the radius of Jupiter, which is a factor of  $10^8$.    Looking at the work of Rott, et al.\cite{Rott:2015nma}, one can see that the bounds on monochromatic neutrinos from the Sun are fairly flat in the $4-10$ GeV mass range, and thus one would expect that, absent evaporation, similar bounds in the $1-4$ GeV mass range would be obtained.   Thus, increasing their bounds by a factor of $10^8$,  DUNE and possibly Hyper-K might be able to set WIMP-nucleon spin-dependent cross-section bounds of roughly  $100$ nanobarns or greater.   There are recent bounds on the spin-dependent cross-section of WIMPs scattering off electrons\cite{Wu:2022jln} which are much smaller than this, but that doesn't necessarily translate into a bound on the WIMP-proton cross-section.     Could such a huge cross-section be possible?   A cross-section this large would be exciting - there could be substantial astrophysical implications since WIMPs would interact more strongly than previously believed (although only in a spin-dependent manner).    In addition,  $\gamma$-emission from final-state radiation could make WIMP dark matter less ``dark".     It would be interesting to study the possibility of a very large spin-dependent WIMP-nucleon cross-section to see if bounds stronger than $100$ nanobarns can be obtained.  If not, DUNE would be able to set the best bounds.

Note added:  After this work was completed, we learned of a recent paper by Leane and Smirnov\cite{Leane:2022hkk} which does a very detailed analysis of the dark matter distribution in the Sun, Earth and Jupiter, disagreeing somewhat with previous analyses.   This will not affect our qualitative results.

\begin{acknowledgments} 
We thank  Danny Marfatia, Jeff Nelson, Tim Tait and especially Mike Kordosky for helpful discussions.   We also thank the NSF for support under Grants PHY-1819575 and PHY-2112460 and GF thanks the Charles Center at William \& Mary and the Virginia Space Grant Consortium for support.
\end{acknowledgments}

%%%%%%%%%%%%%%%%%%%%%%%%%%%%%%%%%%%%%%%%%%%%%%%%%%%%%%%%%%%

\end{document}